\documentclass[journal]{IEEEtran}

\usepackage{amsmath}
\usepackage{amssymb}
\usepackage{latexsym}
\usepackage{multirow}
\usepackage{epsfig}
\usepackage{graphics}
\usepackage{mathrsfs}
\usepackage{algorithmic}
\usepackage{algorithm}

\newcommand{\eqdef}{\stackrel{\triangle}{=}}
\newcommand{\beq}{\begin{equation}}
\newcommand{\enq}{\end{equation}}
\newcommand{\ben}{\begin{eqnarray}}
\newcommand{\enn}{\end{eqnarray}}
\newcommand{\bei}{\begin{itemize}}
\newcommand{\eni}{\end{itemize}}

\newtheorem{theorem}{Theorem}[section]

\begin{document}

\date{}

\title{Integer-Forcing Linear Receiver Design with Slowest Descent Method}

\author{Lili~Wei,~\IEEEmembership{Member,~IEEE}, and Wen Chen,~\IEEEmembership{Senior Member,~IEEE}

\thanks{Manuscript received June 12, 2012; revised January 21, 2013; accepted March 25, 2013. The associate editor coordinating the review of this paper and approving it for publication was Murat Torlak.}
\thanks{L. Wei and W. Chen are with the Department of Electronic Engineering,
Shanghai Jiao Tong University, Shanghai, China 200240. (e-mail: \{liliwei, wenchen\}@sjtu.edu.cn).}
\thanks{This work is supported by the National 973 Project \#2012CB316106 and by NSF China \#61161130529.}
\thanks{Digital Object Identifier XXXXXXXXXX}}

\markboth{IEEE Transactions on Wireless Communications, VOL.~x,
No.~x, xxxxxx~2013} {Shell \MakeLowercase{\textit{WEI and CHEN}}:
Integer-Forcing Linear Receiver Design with Slowest Descent Method}
\maketitle

\begin{abstract}

Compute-and-forward (CPF) strategy is one category of network coding in which a relay will compute and forward a linear combination of source messages according to the observed channel coefficients, based on the algebraic structure of lattice codes. Recently, based on the idea of CPF, integer forcing (IF) linear receiver architecture for MIMO system has been proposed to recover different integer combinations of lattice codewords for further original message detection. In this paper, we consider the problem of IF linear receiver design with respect to the channel conditions. Instead of exhaustive search, we present practical and efficient suboptimal algorithms to design the IF coefficient matrix with full rank such that the total achievable rate is maximized, based on the slowest descent method. Numerical results demonstrate the effectiveness of our proposed algorithms.

\end{abstract}


\IEEEpeerreviewmaketitle

\section{Introduction}

In the past decade, network coding \cite{Cai_NC} has rapidly emerged as a major research area in electrical engineering and computer science. Originally designed for wired networks, network coding is a generalized routing approach that breaks the traditional assumption of simply forwarding data, and allows intermediate nodes to send out functions of their received packets, by which the multicast capacity given by the max-flow min-cut theorem can be achieved. Subsequent works of \cite{Cai_LNC}-\cite{Chou_Poly} made the important observation that, for multicasting, intermediate nodes can simply send out a linear combination of their received packets. Linear network coding with random coefficients is considered in \cite{Medard_Random NC}.  In order to address the broadcast nature of wireless transmission, physical layer network coding \cite{Zhang_PLNC} was proposed to embrace interference in wireless networks in which intermediate nodes attempt to decode the modulo-two sum (XOR) of the transmitted messages. Several network coding realizations in wireless networks are discussed in \cite{Katti_ANC}-\cite{CFNC}.

There is also a large body of works on lattice codes \cite{Lattice1}-\cite{Lattice2} and their applications in communications. For many AWGN networks of interest, nested lattice codes \cite{Nested1} can approach the performance of standard random coding arguments. It has been shown that nested lattice codes (combined with lattice decoding) can achieve the capacity of the point-to-point AWGN channel \cite{Nested2}. Subsequent work of \cite{Lattice3} showed that nested lattice codes achieve the diversity-multiplexing tradeoff of MIMO channel. In the two-way relay networks, a nested lattice based strategy has been developed that the achievable rate is near the optimal upper bound \cite{Lattice-TWRC1}-\cite{Lattice-TWRC3}. The nested lattice codes have the linear structure that ensures that integer combinations of codewords are themselves codewords.

Compute-and-forward (CPF) strategy \cite{CPF}-\cite{Reliable PLNC} is a promising new approach to physical-layer network coding for general wireless networks, beneficial from both network coding and lattice codes. In traditional decode-and-forward (DF) scheme, a relay will decode for individual message and forward; while in compute-and-forward scheme, a relay will compute a linear function of transmitted messages according to the observed channel coefficients. Upon utilizing the algebraic structure of lattice codes, i.e., the integer combination of lattice codewords is still a codeword, the intermediate relay node will compute-and-forward an integer combination of original messages. With enough linear independent equations, the destination can recover the original messages respectively. Subsequent works for design and analysis of the CPF technique have been given in \cite{CPF1}-\cite{CPF5}.

Recently, as a research extension from the idea of CPF strategy, a new linear receiver technique called {\em integer forcing} (IF) receiver for MIMO system has been proposed in \cite{IF1}-\cite{IF2}. In MIMO communication, the destination often utilizes linear receiver architecture to reduce implementation complexity with some performance sacrifice compared with maximum-likelihood (ML) receiver. The standard linear detection methods include zero-forcing (ZF) technique and the minimum mean square error (MMSE) technique \cite{MIMO-book0}. In the newly proposed IF linear receiver, instead of attempting to recover a transmitted codeword directly, each IF decoder recovers a different integer combination of the lattice codewords according to a designed IF coefficient matrix. If the IF coefficient matrix is of full rank, these linear equations can be solved for the original messages.

In this paper, we consider the problem of IF linear receiver design with respect to the channel conditions. Instead of exhaustive search, we present practical and efficient suboptimal algorithms to design the IF coefficient matrix with full rank such that the total achievable rate is maximized, based on the slowest descent method \cite{SDM}. Slowest descent method is a technique to search for discrete points near the continuous-valued slowest descent/ascent lines from the continuous maximizer/minimizer in the Euclidean vector space. This method has been effectively applied to search for binary signatures with quadratic optimization problems in CDMA systems \cite{Wei1}-\cite{Wei2} and MIMO complex discrete signal detection \cite{SDM-MIMO}. In this paper, to design the IF coefficient matrix with integer elements, first we will generate feasible searching set instead of the whole integer searching space based on the slowest descent method. Then we try to pick up integer vectors within our searching set to construct the full rank IF coefficient matrix, while in the meantime, the total achievable rate is maximized.

The notations used in this work are as follows. $\{\cdot\}^T$ denotes the transpose operation, $|\cdot|$ represents the cardinality of a set, $\mathbb Z^n$ denotes the $n$ dimensional integer ring, $\mathbb R^n$ denotes the $n$ dimensional real field. $\mathbb F_p$ denotes a finite field of size p. $\mathbf I_n$ denotes the identity matrix of size $n\times n$, and $\mathbf 0$ denotes the vectors with all zeros elements. $\textit{Re}(\cdot)$ and $\textit{Im}(\cdot)$ denote the real part and the imaginary part. $\det(\cdot)$ denotes the determinant of a matrix. $\partial f/\partial(\mathbf a)$ denotes the partial derivative of function $f$ regarding vector $\mathbf a$. Assume that the $\log$ operation is with respect to base $2$. We use boldface lowercase letters to denote column vectors and boldface uppercase letters to denote matrices.

\section{System Model}

First we note that it is straightforward that a general complex MIMO system $\mathbf y=\mathbf H\mathbf x + \mathbf z$ can be easily converted into an equivalent real system \cite{MIMO-book1} as
\beq
\left[\begin{array}{c}\textit{Re}(\mathbf y)\\ \textit{Im}(\mathbf y)\end{array}\right]=\left[\begin{array}{cc}\textit{Re}(\mathbf H) & -\textit{Im}(\mathbf H)\\ \textit{Im}(\mathbf H) & \textit{Re}(\mathbf H)\end{array}\right] \left[\begin{array}{c}\textit{Re}(\mathbf x)\\ \textit{Im}(\mathbf x)\end{array}\right] + \left[\begin{array}{c}\textit{Re}(\mathbf z)\\ \textit{Im}(\mathbf z)\end{array}\right].\quad
\enq
Hence, we will focus on the real MIMO system for analysis convenience.

We consider the classic MIMO channels with $L$ transmit antennas and $N$ receive antennas. Each transmit antenna delivers an independent data stream which is encoded separately to form the transmitted codewords. We assume that the channel state information is only available at the receiver during each transmission. Let $L=N$ for analysis simplicity.

\begin{figure}
\centerline{\psfig{file=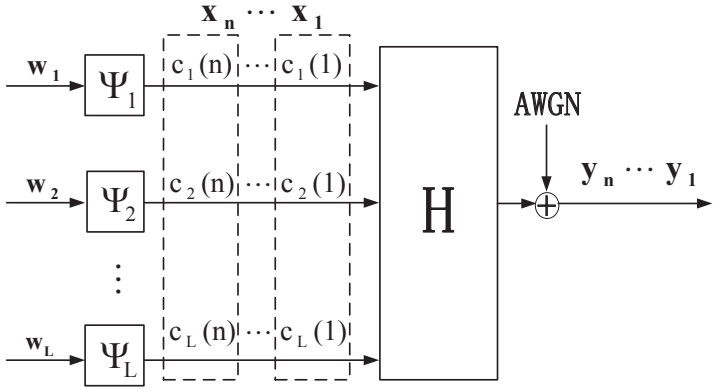,width=3.75in}}
\centering
{\caption{MIMO diagram with independent data streams}}
\end{figure}

Without loss of generality, in one transmission realization, each antenna has a length-$k$ information vectors $\mathbf w_m$ that is drawn independently and uniformly over a prime-size finite field $\mathbb F_p = \{0,1,\cdots,p-1\}$, i.e.,
\beq
\mathbf w_m\in\mathbb F_p^k, \quad m = 1,2,\cdots,L.
\enq

Each antenna is equipped with an encoder $\Psi_m$, that maps the length-$k$ messages $\mathbf w_m$ into the length-$n$ lattice codewords $\mathbf c_m\in \mathbb R^n$,
\beq
\Psi_m: \mathbb F_p^k \to {\mathbb R}^n.
\enq
The codeword satisfies the power constraint of $\frac{1}{n}||\mathbf c_m||^2 \le P$, $m=1,\cdots,L$.

After mapping message $\mathbf w_m$ into a lattice codeword $\mathbf c_m$ with
\beq
\mathbf c_m=[c_m{(1)},c_m{(2)},\cdots,c_m{(n)}]^T,\quad m=1,\cdots,L,
\enq
antenna $m$ will transmit one information codeword $\mathbf c_m$ in one transmission realization with a total of $n$ time slots. In the $i$-th time slot, the transmitted signal vector $\mathbf x_i\in \mathbb R^L$ over $L$ transmit antennas is
\beq
\mathbf x_i = [c_1(i),c_2(i),\cdots,c_L(i)]^T,\quad i=1,\cdots,n.
\enq
The MIMO system diagram with independent data streams is shown in Fig. 1.

Assume a slow fading model where the channel remains constant over the entire codeword transmission. During one transmission realization, at the $i$th time slot, $i=1,\cdots,n$, the received vector $\mathbf y_i\in\mathbb R^L$ is,
\ben
\mathbf y_i = \mathbf H \mathbf x_i + \mathbf z_i,\label{yi}
\enn
where $\mathbf H$ denotes the $L\times L$ channel matrix, with $\mathbf H=[h_{mj}]$ and $h_{mj}$ is the real valued fading channel coefficient from transmit antenna $j$ to receive antenna $m$, and $\mathbf z_i\in \mathbb R^L$ is the additive Gaussian noise. The entries of the channel matrix and the additive noise vector are generated i.i.d. according to a normal distribution $\mathcal{N}(0,1)$.

To facilitate the detection of desired signals from each antenna, in a linear receiver architecture, the receiver will project the received vector ${\mathbf y}_i$ with some matrix $\mathbf B\in \mathbb R^{L\times L}$ to get the effective received vector for further decoding,
\ben
\mathbf d_i = \mathbf B {\mathbf y}_i = \mathbf B {\mathbf H}{\mathbf x}_i + \mathbf B{\mathbf z}_i = \mathbf A{\mathbf x}_i + \mathbf n_i,\label{di}
\enn
where $\mathbf A\eqdef \mathbf B {\mathbf H}$ and $\mathbf n_i\eqdef \mathbf B{\mathbf z}_i$.

The standard suboptimal linear detection methods include the zero-forcing (ZF) receiver and the minimum mean square error (MMSE) receiver,
\ben
\mathbf B_{\textit{ZF}} & = & ({\mathbf H}^T{\mathbf H})^{-1} {\mathbf H}^T,\\
\mathbf B_{\textit{MMSE}} & = & ({\mathbf H}^T{\mathbf H}+\frac{1}{P}\mathbf I_{L})^{-1} {\mathbf H}^T,
\enn
where $(\cdot)^T$ denotes transpose operation and $\mathbf I_{L}$ is the $L\times L$ identity matrix. The ZF technique nullifies the interference such that $\mathbf A_{\textit ZF}=\mathbf I_{L}$ with the effect of noise enhancement. The MMSE receiver maximizes the post-detection signal-to-interference plus noise ratio (SINR) and mitigates the noise enhancement effects. However, both ZF and MMSE receiver have been proved to be largely suboptimal in terms of diversity-multiplexing tradeoff \cite{Linear-receiver}.

We recall the important algebraic structure of lattice codes, that the integer combination of lattice codewords is still a codeword. Instead of restricting matrix $\mathbf A$ to be identity, we may allow ${\mathbf A}$ to be some full rank matrix with integer coefficients, i.e.
\beq
\mathbf A_{\textit{IF}}\in \mathbb {Z}^{L\times L}.
\enq
Then we can first separately recover linear combinations of transmitted lattice codewords with coefficients drawn from matrix ${\mathbf A}_{\textit{IF}}$, which can be easily solved for the original messages. Specifically, if
\beq
\mathbf d_i=[d_1(i),d_2(i),\cdots,d_L(i)]^T,
\enq
then each post-processed output $d_m(i)$ is passed into a separate decoder $\Pi_m$. After one transmission realization with $n$ time slots, at one decoder,
\beq
\Pi_m: \mathbb R^n \to \mathbb F_p^k,
\enq
it maps the post-processed output $d_m(1),d_m(2),\cdots,d_m(n)$ to an estimate $\hat{\mathbf u}_m\in \mathbb F_p^k$ of the linear message combination $\mathbf u_m$, where
\beq
\mathbf u_m = \bigoplus_{l=1}^L q_{ml}\mathbf w_l = \left[\sum_{l=1}^L a_{ml}\mathbf w_l\right] mod\; p,
\enq
where $\bigoplus$ denotes summation over the finite field, $q_{ml}$ is a coefficient taking values in $\mathbb F_p$ and $q_{ml} = a_{ml}\; mod \;p$. The original messages can be recovered from the set of linear equations by a simple inverse operation,
\beq
[\hat{\mathbf w}_1,\hat{\mathbf w}_2,\cdots,\hat{\mathbf w}_L]^T = \mathbf A_{\text{IF}}^{-1}[\hat{\mathbf u}_1,\hat{\mathbf u}_2,\cdots,\hat{\mathbf u}_L]^T.
\enq
The diagram of IF decoder is shown in Fig. 2.

\begin{figure}
\centerline{\psfig{file=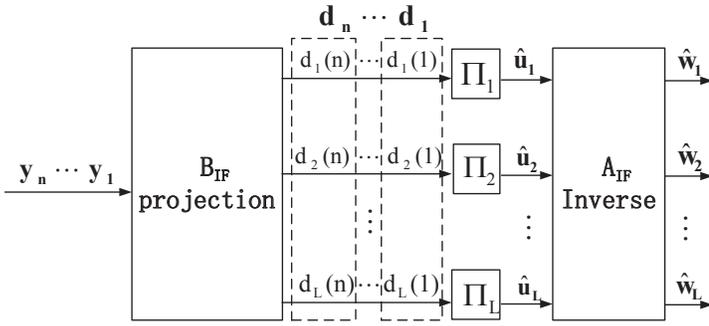,width=3.75in}}
\centering
{\caption{IF decoder diagram}}
\end{figure}

Hence, with integer forcing (IF) technique, the receiver will try to design a projection matrix $\mathbf B_{\textit{IF}}\in \mathbb R^{L\times L}$, such that after the projection process of (\ref{di}), the resulting full rank IF matrix ${\mathbf A_{\textit{IF}}}$ satisfies that $\mathbf A_{\textit{IF}}\in \mathbb {Z}^{L\times L}$ and the achievable rate is maximized. We summarize the results regarding IF receiver in \cite{IF1}-\cite{IF2} in the following theorem.\\

\begin{theorem}\label{theorem-1}
\textit{Let $\mathbf A_{\textit{IF}}=[\mathbf a_1,\mathbf a_2,\cdots,\mathbf a_{L}]^T$ and $\mathbf B_{\textit{IF}}=[\mathbf b_1,\mathbf b_2,\cdots,\mathbf b_{L}]^T$. For each pair of $(\mathbf a_m,\mathbf b_m)$, let ${\textit R}_m$ be the achievable rate of the linear message combination for the $m$-th decoder $\Pi_m$, then
\beq
{\textit R}_m = \frac{1}{2}\log\left(\frac{P}{||\mathbf b_m||^2 + P ||{\mathbf H}^T \mathbf b_m -\mathbf a_m||^2}\right).\label{Rm1}
\enq}

\textit{For a fixed IF coefficient matrix $\mathbf A_{\textit{IF}}$, ${\textit R}_m$ is maximized by choosing\footnote{The optimal projection matrix is \makebox{\scalebox{0.8}{%
\parbox{0.55\textwidth}{$\mathbf B_{\textit{IF}}=\mathbf A_{\textit{IF}}{\mathbf H}^T\left({\mathbf H}{\mathbf H}^T+\frac{1}{P}\mathbf I_{L}\right)^{-1}.$}}}}
\beq
\mathbf b_m^T = \mathbf a_m^T {\mathbf H}^T \left({\mathbf H}{\mathbf H}^T+\frac{1}{P}\mathbf I_{L}\right)^{-1}.\label{bopt}
\enq}
\end{theorem}
\hfill$\square$

According to Theorem \ref{theorem-1}, we plug the optimal $\mathbf b_m$ of (\ref{bopt}) into ${\textit R}_m$ of (\ref{Rm1}), which will result in
\beq
{\textit R}_m = \frac{1}{2}\log\left(\frac{1}{\mathbf a_m^T\, \mathbf Q\, \mathbf a_m}\right),\label{Rm}
\enq
where
\beq
\mathbf Q \eqdef \mathbf I_{L} - {\mathbf H}^T \left({\mathbf H}{\mathbf H}^T + \frac{1}{P}\mathbf I_{L}\right)^{-1}{\mathbf H}.\label{Q}
\enq
The proof of equation (\ref{Rm})-(\ref{Q}) is in Appendix A.

The total achievable rate of the IF receiver is defined as
\ben
{\textit R}_{\textit{total}} & \eqdef & L\,\min \{{\textit R}_1, {\textit R}_2,\cdots, {\textit R}_L\}.
\enn
Hence, the design criteria for optimal IF coeffcient matrix $\mathbf A_{\textit{IF}}$ is
\ben
\mathbf A_{\textit{IF}} & = & \textit{arg}\max_{\mathbf A\in\mathbb Z^{L\times L}}\, {\textit R}_{\textit{total}} \nonumber\\
& = & \textit{arg}\max_{\mathbf A\in\mathbb Z^{L\times L}}\, \min \{{\textit R}_1, {\textit R}_2,\cdots, {\textit R}_L\} \nonumber\\
& = & \textit{arg}\max_{\mathbf A\in\mathbb Z^{L\times L}}\, \min_{m=1,2,\cdots L} \log\left(\frac{1}{\mathbf a_m^T\, \mathbf Q\, \mathbf a_m}\right)\nonumber\\
& = & \textit{arg}\min_{\mathbf A\in\mathbb Z^{L\times L}}\, \max_{m=1,2,\cdots L}\; \mathbf a_m^T\, \mathbf Q \mathbf a_m,\label{A}
\enn
subject to
\beq
\left\{\begin{array}{l}
\mathbf A = [\mathbf a_1, \mathbf a_2, \cdots, \mathbf a_L]^T,\\
\det(\mathbf A) \ne 0,\\
\mathbf a_m \in \mathbb Z^L, \quad\quad m = 1, 2, \cdots, L.
\end{array}\right.\label{A1}
\enq
It means that we need to find integer vectors $\mathbf a_1$, $\mathbf a_2$, $\cdots$, $\mathbf a_{L}$ to construct a full rank matrix $\mathbf A_{\textit{IF}}$, such that the maximum value of $\mathbf a_m^T\, \mathbf Q\, \mathbf a_m$ is minimized.

Solving this optimization problem is critical as it dominates the total achievable rate of the desired IF receiver that sources can reliably communicate with the destination. However, it is the NP-hard quadratic integer programming problem and not convex. The exhaustive search solution is too costly. In this paper, we will propose efficient and practical suboptimal algorithms to design this integer matrix $\mathbf A_{\textit{IF}}$.

\section{Integer Forcing Linear Receiver Design}

To approach the optimization problem of (\ref{A})-(\ref{A1}), first we need to generate some feasible searching set $\Omega\subset \mathbb Z^{L}$ for $\mathbf a_m\in \Omega$, $m = 1,2,\cdots,L$, instead of the whole searching space $\mathbf a_m\in\mathbb Z^{L}$. Then, we will find $L$ linearly independent vectors within this searching set $\Omega$ to construct the IF coefficient matrix $\mathbf A_{\textit{IF}}$.

Accordingly, we propose the following strategy with two steps. In the first step, we generate the searching set $\Omega$ based on slowest descent method \cite{SDM}, which first obtains the optimal minimizer within the continuous domain, and then finds discrete integer points with closest Euclidean distance from slowest ascent lines passing through the optimal continuous minimizer, such that the ``good points'' with small $\mathbf a^{T}\, \mathbf Q\, \mathbf a$ values are within the candidate set $\Omega$.

In the second step, we pick up $\mathbf a_1$, $\mathbf a_2$, $\cdots$, $\mathbf a_L\in \Omega$, to construct the full rank IF coefficient matrix $\mathbf A_{\textit{IF}}=[\mathbf a_1,\mathbf a_2,\cdots,\mathbf a_L]^T$, while in the meantime, the maximum value of $\mathbf a_m^T\, \mathbf Q\, \mathbf a_m$ is minimized. Then, equivalently, this $\mathbf A_{\textit{IF}}$ will maximize the total achievable rate.

\begin{figure}[t]
\centerline{\psfig{file=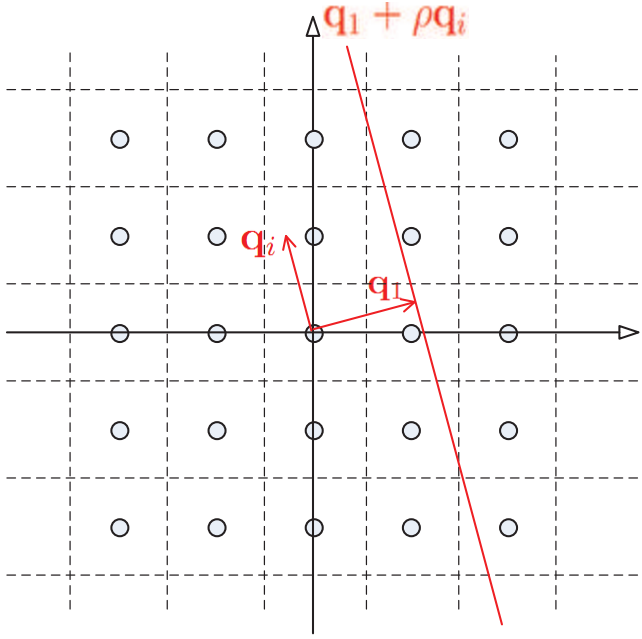,width=3.4in}}
\centering
{\caption{Creation of one slowest ascent line}}
\end{figure}

\subsection{Candidate Set Searching Algorithm with Slowest Descent Method}

We attempt to find the candidate vector set $\Omega$ for $\mathbf a\in\mathbb Z^{L}$, $\mathbf a\ne \mathbf 0$, with small $\mathbf a^{T}\, \mathbf Q\, \mathbf a$ values based on slowest descent method. Note that originally slowest descent method \cite{SDM} is presented for maximization problem with ``slowest descent'' from continuous maximizer. It can be symmetrically applied to minimization problem with ``slowest ascent'' from continuous minimizer. We keep the expression of ``slowest descent method'' for both cases.

First, we relax the constraint of $\mathbf a\in\mathbb Z^{L}$, $\mathbf a\ne \mathbf 0$, and assume, instead, that $\mathbf a$ can be continuous, real-valued $(\mathbf a\in\mathbb R^{L})$ with norm constraint\footnote{The integer constraint $\mathbf a\in\mathbb Z^{L}$, $\mathbf a\ne \mathbf 0$ is equivalent to the constraint $\mathbf a\in\mathbb Z^{L}$, $||\mathbf a||\ge 1$.} $||\mathbf a||\ge 1$, then the corresponding optimization problem becomes,
\beq
\mathbf a_{c}=\textit{arg}\min_{\mathbf a\in\mathbb R^{L},||\mathbf a||\ge 1} \mathbf a^T\mathbf Q\mathbf a.\label{a_real}
\enq

\begin{figure}[t]
\centerline{\psfig{file=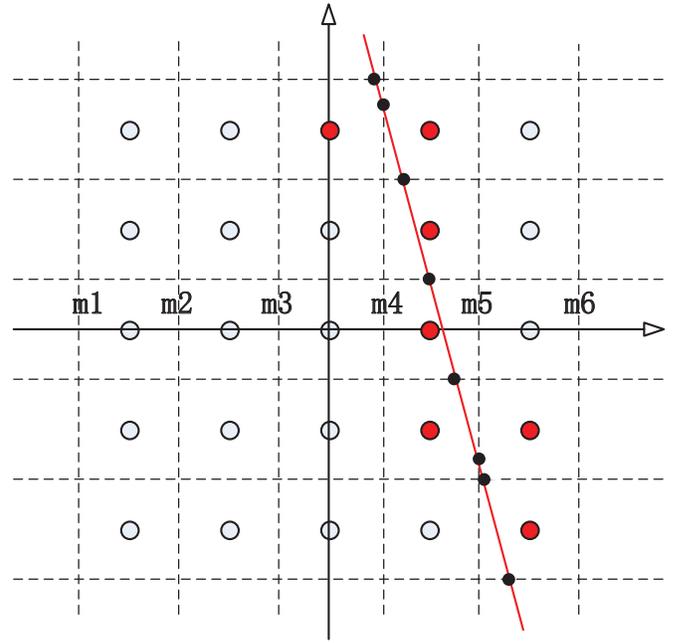,width=3.5in}}
\centering
{\caption{The procedure of slowest descent method}}
\end{figure}

Denote $f(\mathbf a)$ as the cost function, i.e.,
\beq
f(\mathbf a)\eqdef\mathbf a^T\mathbf Q\mathbf a.\label{f}
\enq
Let ${\mathbf g_1,\mathbf g_2,\cdots,\mathbf g_{L}}$ be the $L$ normalized eigenvectors of $\mathbf Q$ with corresponding eigenvalues $\lambda_1\le\lambda_2\le\cdots\le\lambda_{L}$. The real-valued vector for the optimization of (\ref{a_real}) is well known and equal to the eigenvector that corresponds to the minimum eigenvalue of matrix $\mathbf Q$, i.e.,
\beq
\mathbf a_c=\textit{arg}\min_{\mathbf a\in\mathbb R^{2L},||\mathbf a||\ge 1} f(\mathbf a)=\mathbf g_1.\label{q1}
\enq
The Hessian of $f(\mathbf a)$, which is defined as $\mathbf H_{es}^f(\mathbf a)\eqdef(\partial^2 f(\mathbf a))/(\partial\mathbf a\partial \mathbf a^T)$ is
\beq
\mathbf H_{es}^f(\mathbf a)=(\partial^2 f(\mathbf a))/(\partial\mathbf a\partial \mathbf a^T)=\mathbf Q,\label{Hes}
\enq
and well defined at the continuous minimizer vector $\mathbf g_1$.

Then, the eigenvectors $\mathbf g_2$, $\mathbf g_3$, $\cdots$, $\mathbf g_{L}$ of the Hessian $\mathbf H_{es}^f(\mathbf a)=\mathbf Q$ defines mutually orthogonal directions of the least ascent in $f(\mathbf a)$ from the continuous minimizer $\mathbf a_c = \mathbf g_1$ \cite{SDM}. Accordingly, we can construct the slowest ascent lines as
\beq
L(\rho,i)=\mathbf g_1 + \rho \mathbf g_i, \quad \rho\in\mathbb R,\; i=2,\cdots,L,
\enq
which pass through the real minimizer $\mathbf a_c=\mathbf g_1$ with the direction of $\mathbf g_i$.

The creation of one slowest ascent line $\mathbf g_1 + \rho\mathbf g_i$ is shown in Fig. 3. We can see that when $\rho$ takes values from $(-\infty,\infty)$, $\rho\mathbf g_i$ is a line in the direction of $\mathbf g_i$. Hence $\mathbf g_1 + \rho \mathbf g_i$ is a line passing through $\mathbf g_1$ and parallel with the direction of $\mathbf g_i$.

Those ``good'' vectors $\mathbf a\in\mathbb Z^{L}$, $\mathbf a\ne \mathbf 0$ that have small $f(\mathbf a)$ values stays close to those slowest ascent lines. Thus, we are trying to identify those vectors that are closest in Euclidean distance to the above slowest ascent lines, which are the set of candidate vectors
\beq
\Omega = \{\mathbf a(\rho,i): \; \rho\in\mathbb R, i=2,\cdots,L\},
\enq
with
\beq
\mathbf a(\rho,i)=\textit{arg}\min_{\mathbf a\in\mathbb Z^{L},\mathbf a\ne 0}||\mathbf a - L(\rho,i)||.
\enq

Let $M$ be the bound for searching vector $\mathbf a\in\mathbb Z^{L}$ in each coordinate, i.e., each coordinate of $\mathbf a$ takes integer value form $[-M,M]$. Denote $m_j$, $j=1,2,...2M+2$ as the midpoints of adjacent discrete points and the bounds of edge points, i.e., $m_j$ takes value from $[-M-\frac{1}{2},M+\frac{1}{2}]$ with increasing step equal to one. In other words, $m_j\in\Lambda$, where
\beq
\Lambda = \{-M-\frac{3}{2}+j,\; j=1,2,\cdots,2M+2\}.
\enq

For one slowest ascent line $\mathbf g_1+\rho\mathbf g_i$, the closest points to this line changes as $\rho$ varies. To determine those points, it suffices to find the value of $\rho$ for which $\mathbf a(\rho,i)$ exhibits a jump. We may partition the real axis as
\beq
(-\infty, \rho_1],(\rho_1,\rho_2],\cdots,(\rho_{T-1},\rho_T],(\rho_T,+\infty),\label{interval}
\enq
such that within each interval $(\rho_i,\rho_{i+1}]$, $i=1,\cdots,T-1$, there exists exactly one discrete point $\mathbf a$ closest to the slowest ascent line.

Denote $\mathbf g_1=[g_{11},g_{12},\cdots,g_{1,L}]^T$ and $\mathbf g_i=[g_{i1}$, $g_{i2}$, $\cdots$, $g_{i,L}]^T$. Then a jump in one component of $\mathbf a(\rho,i)$ occurs when $\rho$ takes the following value,
\ben
\rho\prime_{(j-1)*L+k} = \frac{m_j-g_{1k}}{g_{ik}}, \label{rho}
\enn
with $k=1,\cdots,L$, and $j=1,\cdots,2M+2$.

After sorting the $\rho\prime_1,\rho\prime_2,\cdots,\rho\prime_{(2M+2)\times 2L}$ calculated from (\ref{rho}) in ascending order, we will get the $\rho_1,\rho_2,\cdots,\rho_T$ for the partition of (\ref{interval}) with $T\le (2M+2)\times L$.

Now that we have those intervals for $\rho$, we are ready to find the points closest to the slowest ascent line $\mathbf g_1+\rho\mathbf g_i$. For one interval $(\rho_j,\rho_{j+1}]$, $j=1,\cdots,T-1$, we can let $\rho$ take the value of
\beq
\rho=\frac{1}{2}(\rho_j+\rho_{j+1}),
\enq
and calculate the closest point $\mathbf a_{(\rho_j,\rho_{j+1}],i}$ by rounding to the closest integer point as
\beq
\mathbf a_{(\rho_j,\rho_{j+1}],i} = \textit{round}\left(\mathbf g_1 + \frac{1}{2}(\rho_j+\rho_{j+1})\mathbf g_i\right).
\enq
After searching through all $T-1$ intervals of $\rho$, we clear the points beyond the bounds $[-M,M]$ in any coordinate, excludes the zero vector\footnote{We add the zero points to facilitate the searching. At the end of the procedure, the vector with all-zero coordinates will exclude.} and put all the final validate points that closest to one slowest ascent line $\mathbf g_1+\rho\mathbf g_i$ in the set of $\Omega_{i-1}$, $i\in\{2,\cdots,L\}$.

Finally, for $J$ slowest ascent lines, $1\le J\le L-1$, we obtain the candidate vector set
\beq
\Omega=\Omega_1\bigcup\Omega_2\bigcup\cdots\bigcup\Omega_{J}.
\enq

The procedure of slowest descent method is shown in Fig. 4. In this example, the bound for searching vector $\mathbf a$ in each coordinate is $M=2$, i.e., each coordinate takes value from $[-2,-1,0,1,2]$. $m_j$, $j=1,2,\cdots,6$ is among $\Lambda=[-2.5,-1.5,-0.5,0.5,1.5,2.5]$. There exists a jump in one component of searching vector $\mathbf a$ when $\rho$ takes value in some interval edge such that the slowest ascent line $\mathbf g_1+\rho\mathbf g_i$ passes the dark solid dots. Within two dark solid dots, there exists only one point closest to the slowest ascent line. After proceeding the described slowest descent method, the outcome candidate points are shown with red solid points.

The complexity reduction of our proposed algorithm for candidate vector set algorithm based on slowest descent method with direct exhaustive search is remarkable. Even when we restrict the direct exhaustive search among the same bound $[-M,M]$ in every coordinate for the searching vector, the direct exhaustive search will need to consider $(2M+1)^{L}$ searching vectors, which is exponential to $L$. While with our proposed algorithm with slowest descent method, we only need to consider $T\le (2M+2)\times L$ points for each slowest ascent line. For $J$ slowest ascent lines, the total vectors considered will be less than $(2M+2)J\times L$. Simulation will show that only a few slowest descent lines need to be considered. Hence, the total complexity reduction of proposed algorithm for candidate vector set algorithm based on slowest descent method is highly favorable.

We summarize our proposed algorithm for candidate vector set $\Omega$ search based on slowest descent method as follows.\\

\vspace{-0.2cm}\hspace{-\parindent}\rule{\linewidth}{1pt}\vspace{-0.0cm}
\vspace{-0.0cm}{\bf Algorithm 1} Candidate Set Search Algorithm with Slowest\\
\hspace*{1.85cm} Descent Method\vspace{-0.2cm}\\
\rule{\linewidth}{.5pt}
{\em Input}: \hspace{0.2cm}Matrix ${\bf Q}$, the bound for each coordinate $M$, and the\\
\hspace*{1.05cm} number of slowest descent lines $J$.\\
{\em Output}: The searching vector candidate set $\Omega$.\\
\hspace{-\parindent}{\underline{Step 1}}: Calculate the eigenvectors $\mathbf g_1$,$\mathbf g_2$,$\cdots$,$\mathbf g_{L}$ of matrix $\mathbf Q$ with corresponding eigenvalues $\lambda_1\le\lambda_2\le\cdots\le\lambda_{L}$.

\hspace{-\parindent}{\underline{Step 2}}: Compute the set of $\Lambda$ which $m_j$, $j=1,2,...2M+2$ takes value from,
\beq
\Lambda = \{-M-\frac{3}{2}+j,\; j=1,2,\cdots,2M+2\}.
\enq

\hspace{-\parindent}{\underline{Step 3}}: For $i=2,\cdots,J+1$ $(1\le J\le L-1)$, do
\bei
\item[{\em (i)}] Obtain the jumping points where $\rho$ takes value as
\beq
\rho\prime_{(j-1)*L+k} = \frac{m_j-g_{1k}}{g_{ik}},\nonumber
\enq
with $k=1,\cdots,L$, $j=1,\cdots,2M+2$.

\item[{\em (ii)}] Sort $\rho\prime_1,\rho\prime_2,\cdots,\rho\prime_{(2M+2)\times L}$ in ascending order into $\rho_1,\rho_2,\cdots,\rho_T$.
\item[{\em (iii)}] For each interval $(\rho_j,\rho_{j+1}]$, $j=1,\cdots,T-1$, we take the value of $\rho$ as
\beq
\rho=\frac{1}{2}(\rho_j+\rho_{j+1}),\nonumber
\enq
and calculate the closest point $\mathbf a_{(\rho_j,\rho_{j+1}],i}$ as
\beq
\mathbf a_{(\rho_j,\rho_{j+1}],i} = \textit{round}\left(\mathbf g_1 + \frac{1}{2}(\rho_j+\rho_{j+1})\mathbf g_i\right).\nonumber
\enq
\item[{\em (iv)}] Clear the points beyond the bounds $[-M,M]$ in any coordinate, excludes the zero vector and put all the final validate points that closest to one slowest ascent line $\mathbf g_1+\rho\mathbf g_i$ in set $\Omega_{i-1}$.
\eni

\hspace{-\parindent}{\underline{Step 3}}: After searching along $J$ slowest ascent lines, we obtain
\beq
\Omega=\Omega_1\bigcup\Omega_2\bigcup\cdots\bigcup\Omega_{J}.
\enq
\rule{\linewidth}{.5pt}

\subsection{Constructing IF Coefficient Matrix $\mathbf A_{\textit{IF}}$}

According to our proposed candidate set search algorithm with slowest descent method, we get the feasible searching set $\Omega$ for IF vectors $\mathbf a_1$, $\mathbf a_2$, $\cdots$, $\mathbf a_L$. With function $f(\cdot)$ defined in (\ref{f}), we sort all vectors within the candidate set $\Omega$ such that
\beq
\Omega=\{\mathbf t_{1}, \mathbf t_{2}, \cdots, \mathbf t_{|\Omega|}: f(\mathbf t_1)\le f(\mathbf t_2)\le \cdots \le f(\mathbf t_{|\Omega|})\}.\label{Set_sort}
\enq

We are trying to choose $L$ linear independent vectors within this sorted set by
\beq
\mathbf a_1=\mathbf t_{i_1}, \mathbf a_2=\mathbf t_{i_2}, \cdots, \mathbf a_L=\mathbf t_{i_{L}}, \;\;\textit{for some}\; i_1<i_2<\cdots<i_{L},
\enq
then the constructed IF coefficient matrix $\mathbf A_{\textit{IF}}$ has full rank $L$. We will select $\mathbf a_1,\mathbf a_2,\cdots,\mathbf a_{L}$ in $\Omega$ based on the greedy search algorithm \cite{CPF4}.

The procedure try to find $\mathbf a_1,\mathbf a_2,\cdots,\mathbf a_{L}$ sequentially and is described as follows,\\
\makebox
{\scalebox{0.93}{%
\parbox{0.5\textwidth}{\ben
\mathbf a_1 & = & \mathbf t_1\nonumber\\
\mathbf a_2 & = & \textit{arg}\min_{\mathbf t\in\Omega}\{f(\mathbf t)\;|\;\mathbf t,\mathbf a_1\; \textit{are linearly independent}\}\nonumber\\
\mathbf a_3 & = & \textit{arg}\min_{\mathbf t\in\Omega}\{f(\mathbf t)\;|\;\mathbf t,\mathbf a_1,\mathbf a_2\; \textit{are linearly independent}\}\nonumber\\
 & \vdots & \nonumber\\
\mathbf a_L & = & \textit{arg}\min_{\mathbf t\in\Omega}\{f(\mathbf t)\;|\;\mathbf t,\mathbf a_1,\cdots,\mathbf a_{L-1}\; \textit{are linearly independent}\}\nonumber
\enn}}}

We summarize this procedure to constructing the full rank optimal matrix $\mathbf A_{\textit{IF}}$ with candidate vector set $\Omega$ as follows.\\


\vspace{-0.2cm}\hspace{-\parindent}\rule{\linewidth}{1pt}\vspace{-0.0cm}
\vspace{-0.0cm}{\bf Algorithm 2} IF Coefficient Matrix Constructing with Greedy\\
\hspace*{1.85cm} Search Algorithm\vspace{-0.2cm}\\
\rule{\linewidth}{.5pt}
{\em Input}:\hspace{0.2cm} Searching set $\Omega$, Matrix $\mathbf Q$.\\
{\em Output}: The IF coefficient matrix $\mathbf A_{\textit{IF}}$ with full rank that gives\\
\hspace*{1.1cm} the maximum total achievable rate ${\textit R}_{\textit{total}}$.\\
\hspace{-\parindent}{\underline{Step 1}}: Let $f(\mathbf t)=\mathbf t^T \mathbf Q \mathbf t$ and sort all vectors in the searching candidate set $\Omega$ such that
\beq
\Omega=\{\mathbf t_{1}, \mathbf t_{2}, \cdots, \mathbf t_{|\Omega|}: f(\mathbf t_1)\le f(\mathbf t_2)\le \cdots \le f(\mathbf t_{|\Omega|})\}.\nonumber
\enq
Set $\mathbf a_1=\mathbf t_1$. Initialize $i=1$ and $j=1$.

\hspace{-\parindent}{\underline{Step 2}}: While $i<|\Omega|$ and $j<L$, do
\bei
\item[{\em (i)}] Set $i=i+1$.
\item[{\em (ii)}] Check whether $\mathbf t_i$, $\mathbf a_1$,$\cdots$,$\mathbf a_j$ are linearly independent according to the determinant of the Gramian matrix. Let $\mathbf G = [\mathbf t_i, \mathbf a_1, \cdots, \mathbf a_j]$. If $\det(\mathbf G^T \mathbf G) \ne 0$, then $j=j+1$ and $\mathbf a_j=\mathbf t_i$. Else goto {\underline{Step 2}}.
\eni

\hspace{-\parindent}{\underline{Step 3}}: Construct the full rank IF coefficient matrix $\mathbf A$ as $\mathbf A=[\mathbf a_1,\mathbf a_2,\cdots,\mathbf a_L]^T$.\\
\rule{\linewidth}{.5pt}

\section{Experimental Studies}

We present numerical results to evaluate the performance of our proposed algorithm for IF receiver design. First, we discuss the IF performance with respect to some parameters. One of them is $J$, which represents how many slowest ascent lines are chosen during the candidate set search procedure. For an example setting with $L=8$, which $J$ can take values in $[1,2,\cdots,7]$, we average over 10000 randomly generated channel realizations and show the average rate in Fig. 5. The bound for each coordinate of searching vector is $M=2$. We can see that as $J$ increases, since the candidate set $\Omega=\Omega_1\bigcup\Omega_2\cdots\bigcup\Omega_{J}$ will expand, the performance is getting better as expected. However, we do not need to span all slowest ascent lines since the performance improvements are becoming indifferent as $J$ becomes larger, for example, $J\ge 4$, which means further increase of $J$ will not have much effect on the performance.

\begin{figure}[!b]
\centerline{\psfig{file=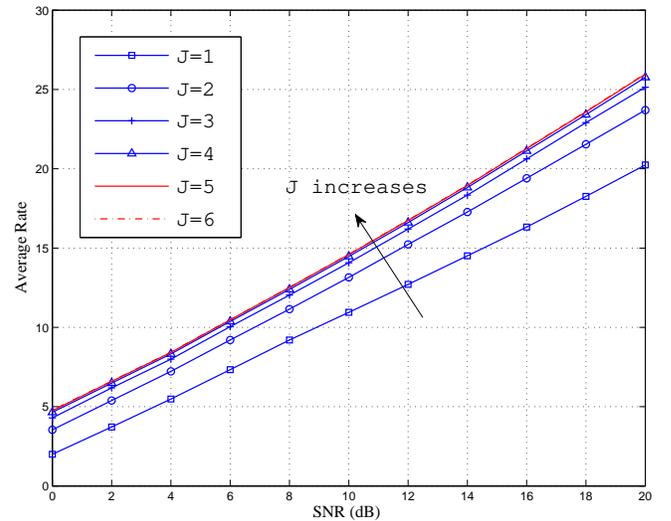,width=4in}}
\centering
{\caption{The performance of IF receiver design regarding $J$ ($L=8$, $M=2$)}}
\end{figure}

\begin{figure}[!b]
\centerline{\psfig{file=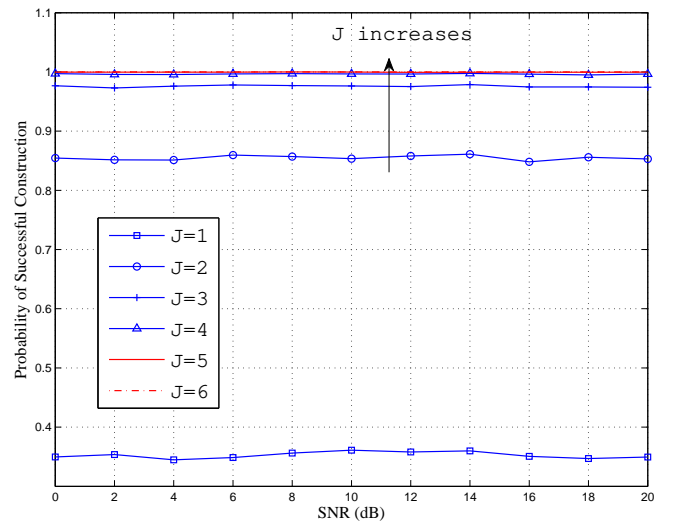,width=4in}}
\centering
{\caption{The probability of successful construction of IF matrix regarding $J$ ($L=8$,$M=2$)}}
\end{figure}

\begin{figure}[!t]
\centerline{\psfig{file=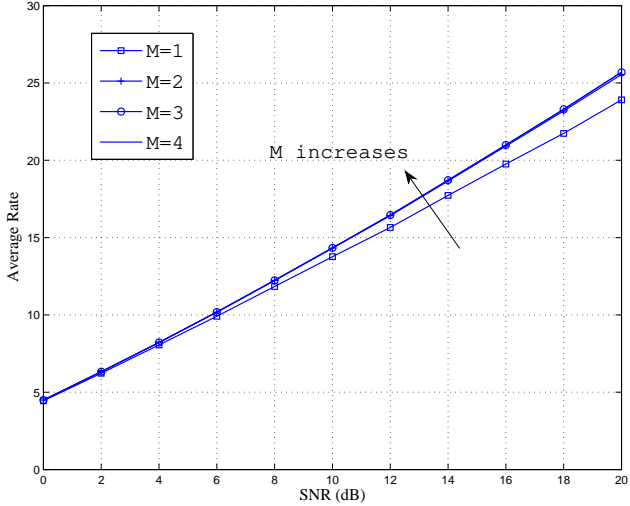,width=3.9in}}
\centering
{\caption{The performance of IF receiver design regarding $M$($L=8$,$J=4$)}}
\end{figure}

\begin{figure}[h]
\centerline{\psfig{file=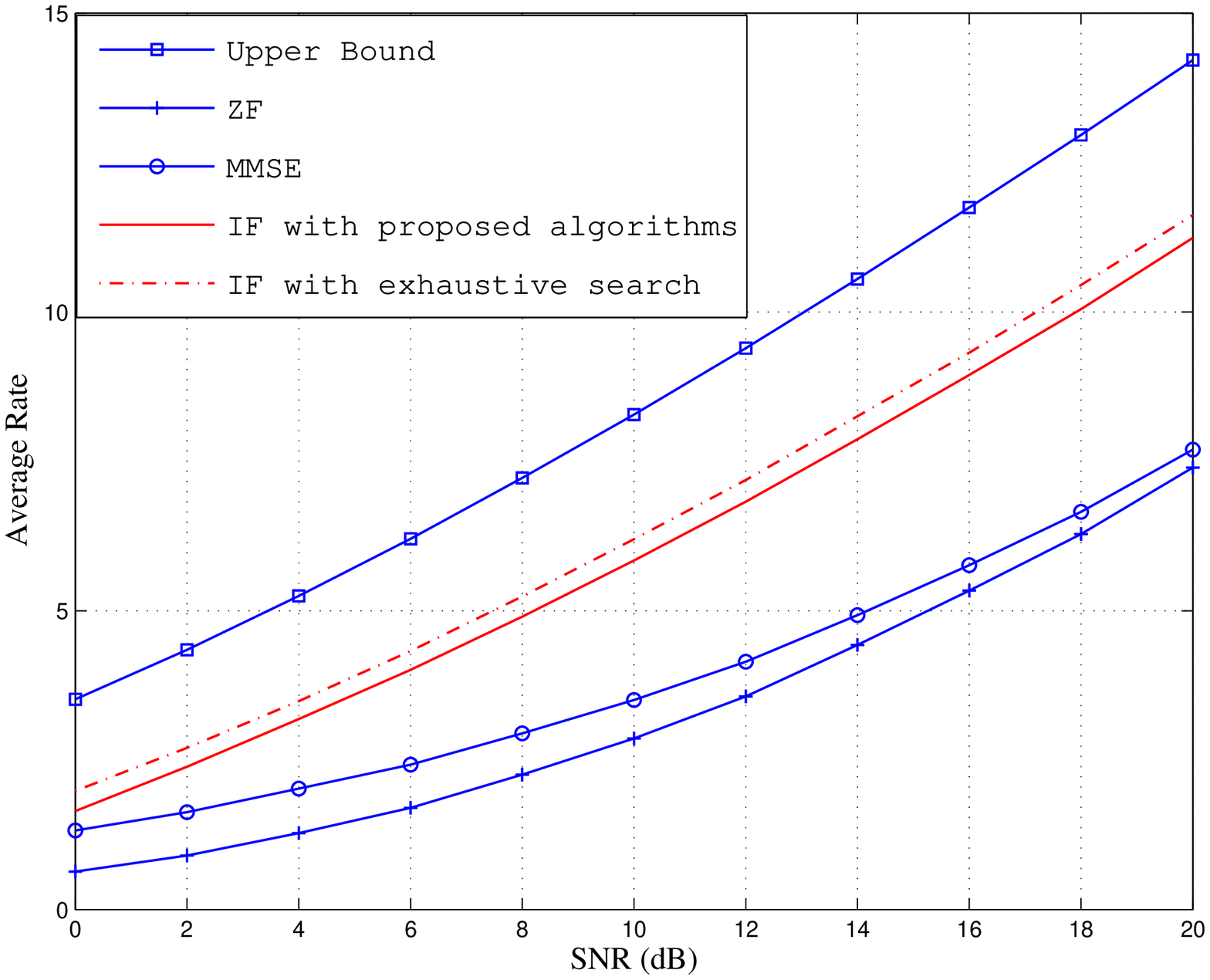,width=3.9in}}
\centering
{\caption{The comparisons of different linear detectors($L=4$,$J=2$,$M=2$)}}
\end{figure}

\begin{figure}[!t]
\centerline{\psfig{file=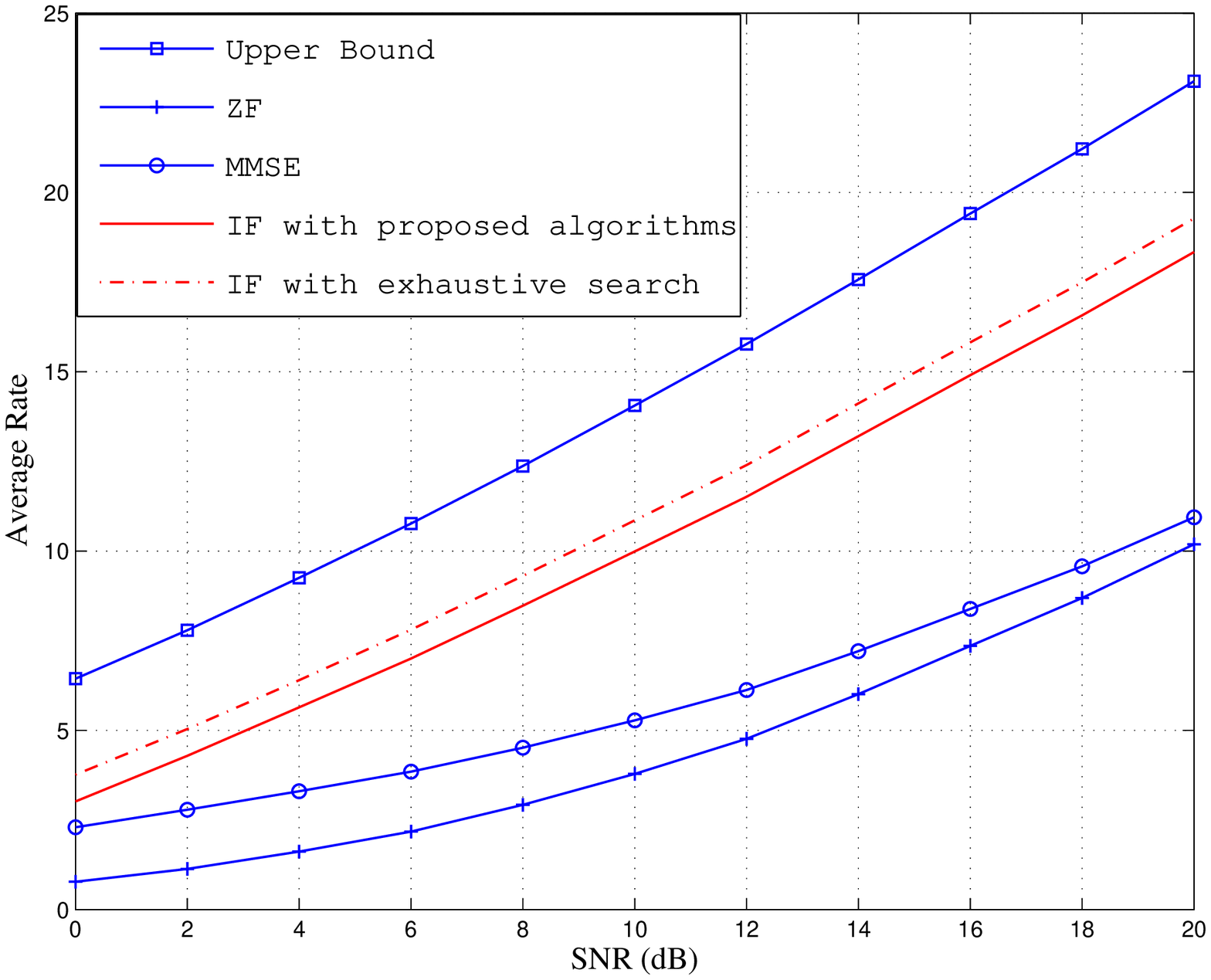,width=3.9in}}
\centering
{\caption{The comparisons of different linear detectors($L=6$,$J=3$,$M=2$)}}
\end{figure}

\begin{figure}[h]
\centerline{\psfig{file=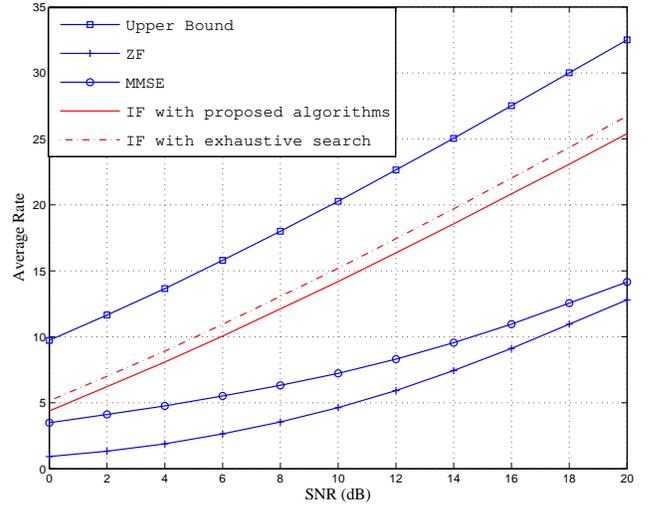,width=3.9in}}
\centering
{\caption{The comparisons of different linear detectors($L=8$,$J=4$,$M=2$)}}
\end{figure}

With the same simulation settings, in Fig. 6, we show the probability of successful construction of IF matrix after running the proposed candidate set search algorithm with slowest descent method and IF coefficient matrix constructing with greedy search algorithm. When the candidate set $\Omega$ is too small, there are chances that we could not construct a full rank IF coefficient matrix with those vectors in $\Omega$. Since the candidate set $\Omega$ is expanding with regarding to the setting of $J$, the successful construction probability will raise as $J$ increases, which can be observed in Fig. 6. On the other hand, when $J\ge 4$, the successful probability of IF integer matrix is almost equal to 1, which means that we do not need to consider all $L-1$ slowest ascent lines.

Next, we investigate the IF performance regarding the parameter of $M$, which is the bound for each coordinate of searching vector. In other words, for searching vector $\mathbf a=[a_1,a_2,\cdots,a_{L}]$, the bound $M$ restrict $a_i\in \mathbb Z$ and $|a_i|\le M$. Obviously when $M$ is expanding, the IF performance is getting improved since we are searching among a broader space. In Fig. 7, we show the average rate with respect to the setting example of $M$ with $L=8$ and $J=4$ over 10000 randomly generated channel realizations. $M=1$ means each coordinate of the searching vector takes value within $[-1,0,1]$; $M=2$ means each coordinate of the searching vector takes value within $[-2,-1,0,1,2]$, and so on. We can observe that when $M\ge 2$, further expanding of $M$ actually does not help improving the performance any more. This is because that those ``good'' vectors $\mathbf a\in\mathbb Z^{L}$, $\mathbf a\ne \mathbf 0$ that have small $\mathbf a^T\mathbf Q\mathbf a$ values stay close to the continuous minimizer (\ref{q1}) and tends to have small coordinates. As shown in Fig. 7, $M=2$ is good enough for this setting example. Hence, for different settings, we can investigate and decide the proper $J$ and $M$ parameters for further realizations.

After the parameters setting discussion, we are ready to compare the performance of different linear detection methods. The standard linear detection methods of ZF and MMSE are
included for comparisons. We also take account in the channel capacity of $C = \frac{1}{2}\log\det(\mathbf I_{L} + P \mathbf H \mathbf H^T)$, which represents the upper bound
for all receiver structures, linear or non-linear including joint ML. We include the theoretical brute forth exhaustive search to construct candidate set for IF receiver design as well, which has high complexity and not applicable in practise. In Fig. 8, we show the rate comparisons over MIMO channels with $L=4$, $J=2$, $M=2$ and average of 10000
randomly generated channel realizations. The average rate of the designed IF linear receiver with IF coefficient matrix $\mathbf A_{\textit{IF}}$ obtained by our proposed algorithms, is significantly improved compared to ZF and MMSE linear receiver. In addition, the performance of our proposed SDM based algorithms approaches the exhaustive search results. As the complexity reduction analysis in Section III. A, our proposed algorithms are practical and efficient. We repeat our experimental in Fig. 9 with $L=6$, $J=3$, $M=2$,  in Fig. 10 with $L=8$, $J=4$, $M=2$. Similar conclusions can be drawn.

\section{Conclusions}

Motivated by recently presented integer-forcing linear receiver architecture, in this paper, we consider the problem of IF linear receiver design with respect to the channel conditions. We present practical and efficient algorithms to design the IF full rank coefficient matrix with integer elements, such that the total achievable rate is maximized, based on the slowest descent method.  First we will generate feasible searching set with integer vector search near the continuous-valued lines of least metric increase from the continuous minimizer in the Euclidean vector space. Then we try to pick up integer vectors within our searching set to construct the full rank IF coefficient matrix, while in the meantime, the total achievable rate is maximized. Numerical studies discuss the parameter settings and comparisons of other traditional linear receivers.

\vspace{0.3cm}

{\begin{center}APPENDIX\end{center}}

\vspace{0.1cm}

{\begin{center}A. Proof of Equations (\ref{Rm})-(\ref{Q})\end{center}}
\vspace{0.3cm}

\hspace{-\parindent}Denote $\mathcal L\eqdef\frac{1}{P}||\mathbf b_m||^2 + ||{\mathbf H}^T \mathbf b_m -\mathbf a_m||^2$, then
\beq
\mathcal L = \frac{1}{P}\mathbf b_m^T\mathbf b_m + \mathbf b_m^T{\mathbf H}{\mathbf H}^T\mathbf b_m - \mathbf a_m^T{\mathbf H}^T\mathbf b_m - \mathbf b_m^T{\mathbf H}\mathbf a_m+\mathbf a_m^T\mathbf a_m.\nonumber
\enq

\hspace{-\parindent}Let $\mathcal F\eqdef{\mathbf H}{\mathbf H}^T+\frac{1}{P}\mathbf I_{L}$. Accordingly the optimal $\mathbf b_m^T$ can be written as
\ben
\mathbf b_m^T & = & \mathbf a_m^T {\mathbf H}^T \left({\mathbf H}{\mathbf H}^T+\frac{1}{P}\mathbf I_{L}\right)^{-1}\nonumber\\
& = & \mathbf a_m^T{\mathbf H}^T\mathcal F^{-1}.\nonumber
\enn

\hspace{-\parindent}Plug in this optimal $\mathbf b_m^T$ in $\mathcal L$, we have
\ben
\mathcal L & = & \mathbf a_m^T\frac{{\mathbf H}^T\mathcal F^{-1}}{P}\mathbf b_m + \mathbf a_m^T{\mathbf H}^T\mathcal F^{-1}{\mathbf H}{\mathbf H}^T\mathbf b_m - \mathbf a_m^T{\mathbf H}^T\mathbf b_m\nonumber\\
 & & - \mathbf a_m^T{\mathbf H}^T\mathcal F^{-1}{\mathbf H}\mathbf a_m + \mathbf a_m^T\mathbf a_m\nonumber\\
& = & \mathbf a_m^T{\mathbf H}^T\mathcal F^{-1}\underbrace{\left(\frac{1}{P}\mathbf I_{L}+{\mathbf H}{\mathbf H}^T\right)}_{\mathcal F}\mathbf b_m - \mathbf a_m^T{\mathbf H}^T\mathbf b_m\nonumber\\
 & & - \mathbf a_m^T{\mathbf H}^T\mathcal F^{-1}{\mathbf H}\mathbf a_m + \mathbf a_m^T\mathbf a_m\nonumber\\
& = & \mathbf a_m^T\mathbf a_m - \mathbf a_m^T{\mathbf H}^T\mathcal F^{-1}{\mathbf H}\mathbf a_m\nonumber\\
& = & \mathbf a_m^T\underbrace{\left(\mathbf I_{L}-{\mathbf H}^T\mathcal F^{-1}{\mathbf H}\right)}_{\mathbf Q}\mathbf a_m.\nonumber \enn

\hspace{-\parindent}Hence,
\ben
{\textit R}_m & = & \frac{1}{2}\log\left(\frac{P}{||\mathbf b_m||^2 + P ||{\mathbf H}^T \mathbf b_m -\mathbf a_m||^2}\right)\nonumber\\
& = & \frac{1}{2}\log\left(\frac{1}{\mathcal L}\right)\nonumber\\
& = & \frac{1}{2}\log\left(\frac{1}{\mathbf a_m^T\mathbf Q\mathbf a_m}\right),\nonumber
\enn
where
\ben
\mathbf Q & = & \mathbf I_{L}-{\mathbf H}^T\mathcal F^{-1}{\mathbf H}\nonumber\\
& = & \mathbf I_{L} - {\mathbf H}^T \left({\mathbf H}{\mathbf H}^T + \frac{1}{P}\mathbf I_{L}\right)^{-1}{\mathbf H}.\nonumber
\enn
\hfill$\square$


\begin{thebibliography}{99}


\bibitem{Cai_NC}
R. Ahlswede, N. Cai, S. R. Li and R. W. Yeung, ``Network information flow'', {\em IEEE Trans. Info. Theory}, vol. 46, no. 4, pp. 1204-1216, July 2000.

\bibitem{Cai_LNC}
S. R. Li, R. Ahlswede and N. Cai, ``Linear network coding'', {\em IEEE Trans. Info. Theory}, vol. 49, no. 2, pp. 371-381, Feb. 2003.

\bibitem{Medard_Algebraic NC}
R. Koetter and M. Medard, ``An algebraic approach to network coding'', {\em IEEE/ACM Trans. Networking}, vol. 11, no. 5, pp.782-795, Oct. 2003.

\bibitem{Chou_Poly}
S. Jaggi, P. Sanders, P. A. Chou, M. Effros, S. Egner, K Jain and L. Tolhuizen, ``Polynomial time algorithms for multicast network code construction'', {\em IEEE Trans. Info. Theory}, vol. 51, no.6, June. 2005.

\bibitem{Medard_Random NC}
T. Ho, R. Koetter, M. Medard, D. R. Karger, M. Effros, J. Shi and B. Leong, ``A random linear network coding approach to multicast'', {\em IEEE Trans. Info.  Theory}, vol. 52, no. 10, pp. 4413-4430, Oct. 2006.


\bibitem{Zhang_PLNC}
S. Zhang, S. Liew and P. P. Lam, ``Hot topic: Physical-layer network coding'', in {\em Proc. 12th Ann. ACM Int. Conf. on Mobile Comput. $\&$ Netw.}, pp. 358-365, Los Angeles, USA, Sept. 2006.

\bibitem{Katti_ANC}
S. Katti, S. Gollakota, and D. Katabi, ``Embracing wireless interference: Analog network coding'', in {\em Proc. ACM SIGCOMM}, Kyoto, Japan, pp. 397-408, Aug. 2007.

\bibitem{Katti_XOR}
S. Katti, H. Rahul, W. Hu, D. Katabi, M. Medard and J. Crowcroft, ``XORs in the air: Practical wireless network coding'', in {\em Proc. ACM SIGCOMM}, Pisa, Italy, pp. 243-254, Sept. 2006.

\bibitem{Wu_Exchange}
Y. Wu, P. A. Chou and S. Kung, ``Information exchange in wireless networks with network coding and physical-layer broadcast'', Microsoft Research, Redmond, WA, Tech. Rep. MSR-TR-2004-78, Aug. 2004.

\bibitem{Medard-wireless}
C. Fragouli, D. Katabi, A. Markopoulou, M. Medard and H. Rahul, ``Wireless network coding: Opportunities $\&$ Challenges'', in {\em Proc. IEEE MILCOM}, Orlando, FL, Oct. 2007.

\bibitem{CFNC}
T. Wang and G. B. Giannakis, ``Complex field network coding for multiuser cooperative communications,¡± {\em IEEE J. Sel. Areas Commun.}, vol. 26, no. 2, pp. 561-571, Apr. 2008.



\bibitem{Lattice1}
R. Zamir, ``Lattices are everywhere'', in {\em Proceedings of the 4th Annual Workshop on Info. Theory and its Applications}, La Jolla, CA, Feb. 2009.

\bibitem{Lattice2}
F. Oggier and E. Viterbo, ``Algebraic number theory and code design for rayleigh fading channels'', in {\em Foundations and Trends in Commun. and Info. Theory}, vol. 1, pp. 333-415, 2004.

\bibitem{Nested1}
R. Zamir, S. Shamai and U. Erez, ``Nested linear/lattice codes for structed multiterminal binning'', {\em IEEE Trans. Info. Theory}, vol. 48, no. 6, pp. 1250-1276, June 2002.

\bibitem{Nested2}
U. Erez and R. Zamir, ``Achieving (1/2)log(1+SNR) on the AWGN channel with lattice encoding and decoding'', {\em IEEE Trans. Info. Theory}, vol. 50, no. 10, pp. 2293-2314, Oct. 2004.

\bibitem{Lattice3}
H. E. Gamal, G. Caire and M. O. Damen, ``Lattice coding and decoding achieve the optimal diversity-multiplexing tradeoff of MIMO channels'', {\em IEEE Trans. Info. Theory}, vol. 50, pp. 968-985, June 2004.


\bibitem{Lattice-TWRC1}
M. Wilson, K. Narayanan, H. Pfister and A. Sprintson, ``Joint physical layer coding and network coding for bidirectional relaying,'' {\em IEEE Trans. Info. Theory}, vol. 56, pp. 5641-5654, Nov. 2010.

\bibitem{Lattice-TWRC2}
B. Hassibi, A. Sezgin, M. Khajehnejad and A. Avestimehr, ``Approximate capacity region of the two-pair bidirectional gaussian relay network,'' in {\em IEEE Inter. Symp. Info. Theory}, pp. 2018-2022, Seoul, Korea, June, 2009.

\bibitem{Lattice-TWRC3}
I. Baik and S. Y. Chung, ``Network coding for two-way relay channels using lattices,'' in {\em IEEE Inter. Conf. Commun.}, pp. 3898-3902, Beijing, China, May 2008.


\bibitem{CPF}
B. Nazer and M. Gastpar, ``Compute-and-forward: harnessing interference through structured codes'', {\em IEEE Trans. Info. Theory}, vol. 57, no. 10, pp. 6463-6486, Oct. 2011.

\bibitem{Reliable PLNC}
B. Nazer and M. Gastpar, ``Reliable physical layer network coding'', {\em IEEE Proceedings}, vol. 99, no. 3, pp. 438-460, Mar. 2011.


\bibitem{CPF1}
M. Nokleby and B. Aazhang, ``Lattice coding over the relay channel``, {\em IEEE Inter. Conf. Commun.}, Kyoto, Japan, June 2011.

\bibitem{CPF2}
C. Feng, D. Silva and F. R. Kschischang, ``An algebraic approach to physical-layer network coding'', {\em submitted to IEEE Trans. Info. Theory}, available online: http://arxiv.org/abs/1108.1695.

\bibitem{CPF3}
J. C. Belfiore and M. A. V. Castro, ``Managing interferece through space-time codes, lattice reduction and network coding'', in {\em proc. IEEE Info. Thoery Workshop}, Dublin, Ireland, Aug. 2010.

\bibitem{CPF4}
L. Wei and W. Chen, ``Compute-and-forward network coding design over multi-source multi-relay channels'', {\em IEEE Trans. Wireless Communications}, vol. 11, no. 9, pp. 3348-3357, Sept. 2012.

\bibitem{CPF5}
L. Wei and W. Chen, ``Efficient compute-and-forward network codes search for two-way relay channel'', {\em IEEE Commun. Letters}, vol. 16, no. 8, pp. 1204-1207, Aug. 2012.




\bibitem{IF1}
J. Zhan, B. Nazer, U. Erez and M. Gastpar, ``Integer-forcing linear receivers'', in {\em IEEE Inter. Symp. Info. Theory}, pp. 1022-1026, Austin, Texas, June 2010.

\bibitem{IF2}
J. Zhan, B. Nazer, U. Erez and M. Gastpar, ``Integer-forcing linear receivers: a new low-complexity MIMO architecture'', in {\em IEEE Veh. Tech. Conf.}, Ottawa, Canada, Sept. 2010.

\bibitem{MIMO-book0}
D. Tse and P. Viswanath, {\em Fundamentals of wireless communications}. New York, NY, USA: Cambridge University Press, 2005.

\bibitem{SDM}
P. Spasojevic and C. N. Georghiades, ``The slowest descent method and its application to sequence estimation'', {\em IEEE Trans. Commun.}, vol. 49, no. 9, pp. 1592-1604, Sept. 2001.

\bibitem{Wei1}
L. Wei, S. N. Batalama, D. A. Pados and B. W. Suter, ``Adapative binary signature design for code-division multiplexing'', {\em IEEE Trans. Wireless Commun.}, vol. 7, no. 7, pp. 2798-2804, July, 2008.

\bibitem{Wei2}
L. Wei, S. N. Batalama, D. A. Pados and B. W. Suter, ``Upward scaling of minimum-TSC binary signature sets'', {\em IEEE Commun. Lett.}, vol. 11, no. 11, pp. 889-891, Nov. 2007.

\bibitem{SDM-MIMO}
F. Simoens, D. V. Welden, H. Wymeersch and M. Moeneclaey, ``Low complexity MIMO detection based on the slowest descent method'', {\em IEEE Commun. Lett.}, vol. 11, no. 5, pp. 429-431, May 2007.


\bibitem{MIMO-book1}
Y. S. Cho, J. Kim, W. Y. Yang and C. G. Kang, ``MIMO-OFDM wireless communications with matlab'', John Wiley \& Sons Pte Ltd, ISBN: 978-0-470-82561-7, 2010.


\bibitem{Linear-receiver}
K. R. Kumar, G. Caire, and A. L. Moustakas, ``Asymptotic performance of linar receivers in MIMO fading channels'', {\em IEEE Trans. Info. Theory}, vol. 55, no. 10, pp. 4398-4418, Oct. 2009.



\end{thebibliography}
\end{document}